\documentclass[aps,prl,showpacs,twocolumn]{revtex4}
\usepackage[usenames]{color}
\newcommand{\comment}[1]{}
\usepackage{graphicx}
\usepackage{epsfig}
\usepackage{amsmath}

\usepackage{color}

\newcommand{\map}[1]{{\color{black}{#1}}}

\begin{document}

\title{Vortex Lattice Locking in Rotating Two-Component 
Bose-Einstein Condensates}

\author{Ryan Barnett$^{1}$, Gil Refael$^{1}$, 
Mason A. Porter$^{2}$, 
and Hans Peter B\"uchler$^{3}$}

\affiliation{$^{1}$Department of Physics, California Institute 
of Technology, MC 114-36, Pasadena, California 91125}
\affiliation{$^2$Oxford Centre for Industrial and Applied 
Mathematics, Mathematical Institute, University of Oxford, OX1 3LB, UK}
\affiliation{$^3$ Institut f\"ur Theoretische 
Physik III, Universit\"at Stuttgart, 70550}

\date{\today}

\begin{abstract}
The vortex density of a rotating superfluid, divided by its particle
mass, dictates the superfluid's angular velocity through the Feynman
relation. To find how the Feynman relation applies to superfluid mixtures, we
investigate a rotating two-component 
Bose-Einstein condensate, composed of bosons with different masses. We
find that in the case of sufficiently strong interspecies attraction,
the vortex lattices of the two condensates lock and rotate at the drive
frequency, while the superfluids themselves rotate at two different
velocities, whose ratio is the ratio between the particle mass of the two
species. In this paper, we characterize the vortex-locked state, 
establish its regime of stability, and find
that it surives within a disk smaller than a critical radius, beyond which
vortices become unbound, and the two Bose-gas rings  rotate together at
the frequency of the external drive.  
\end{abstract}


\maketitle


After the first experimental realization of 
Bose-Einstein condensates (BECs) of
alkali atoms, their study has experienced enormous 
advancements \cite{pethick02}.  Among the major 
threads of investigation in BECs has been the study of vortices both experimentally \cite{matthews99, madison00, abo-shaeer01} and theoretically \cite{fetter01}.
It is known from the classic works of Onsager and 
Feynman \cite{onsager49,
feynman55} that superfluids rotate by nucleating vortices.  
\map{W}hen there are several vortices present, they 
form a triangular Abrikosov vortex
lattice \cite{abrikosov57}, with density given by
\begin{equation}
	\label{Eq:feynman}
	\rho_v = \frac{m \Omega}{\pi \hbar}\,,
\end{equation}
where $m$ is the mass of \map{a constituent boson} and $\Omega$ is the rate at
which the superfluid \map{-- which rotates with the vortex lattice --} is being rotated (see, for instance, \map{Ref.~}\cite{donnelly91})\map{.}
The so-called ``Feynman relation'' (\ref{Eq:feynman}) states that, 
on average, a
uniform superfluid rotates like a rigid body.  It has been shown that
corrections to Eq.~(\ref{Eq:feynman}) due to the typical experimental situation
of nonuniform superfluid density resulting from a harmonic trap are small
\cite{sheehy04} but experimentally observable \cite{coddington04}.
Vortex physics becomes much more intriguing in multi-component BEC's. 
So far, the investigation of vortex lattices in
multi\map{-}component BECs utilized different hyperfine levels of the
constituent atoms to obtain multi-component condensates (e.g.,
\cite{mueller02,kasamatsu05a}). Thus the mass of all condensate
components was identical, and the generalization of
Eq.~(\ref{Eq:feynman}) straightforward.

\begin{figure}
\includegraphics[width=3.4in]{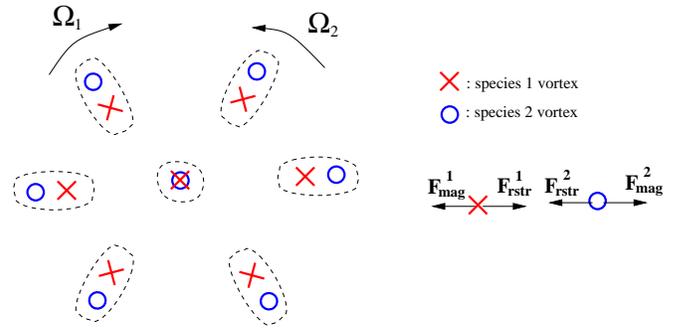}
\caption{Schematic diagram of bound vortex pairs (\map{for species 
with masses $m_1$ and $m_2< m_1$}) 
in the rest frame of the vortex lattice in the limit of large 
interspecies interaction
in which the two vortex lattices are locked.  The Magnus force
${\bf F}_{\rm mag}$ opposes such locking and is balanced by a restoring force
${\bf F}_{\rm rstr}$ due to the interspecies vortex-vortex interaction. }
\label{Fig:cartoon}
\end{figure}

In this paper we ask  what are the consequences of the Feynman condition,
Eq. (\ref{Eq:feynman}), in a system of interacting two-component
rotating condensates with {\it different masses}. We find that for sufficiently large
attractive interactions, the Feynman condition leads to a novel state.
The two components, rather than rotate together at the drive frequency,
rotate at angular velocities $\Omega_{1,2}$ inversely proportional to their
masses, $m_{1,2}$, such that: 
\begin{equation}
m_1 \Omega_1=m_2\Omega_2
\label{results1}
\end{equation}
while the vortex lattices of the two components lock at the drive
angular velocity $\Omega_v$, lying between $\Omega_1$ and $\Omega_2$
(see Fig.~\ref{Fig:cartoon}). 
Qualitatively, the attractive interspecies interaction leads to an
attraction between vortices of the two flavors. If it is
sufficiently strong, vortices pair, and the lowest\map{-}energy vortex
configuration then occurs when the vortex lattices of the two flavors are ``locked''
together, rotate at the same rate, and have
essentially the same density. Eq.~\map{(}\ref{Eq:feynman}\map{)}
then reads: 
\begin{equation}
\rho_v^{1}=\frac{m_1 \Omega_1}{\pi \hbar}\approx\frac{m_2 \Omega_2}{\pi \hbar}=\rho_v^{2}
\end{equation}
where $\rho_v^{1,2}$ are the vortex densities of the two flavors. This state is strongly related to experiments in
Ref. \onlinecite{tung06}, where a vortex lattice was locked to an
optical lattice with a similar periodicity. 

As we show below, this state survives within a finite disk about the
center of the rotating condensate. The relative motion between the
vortices and the condensate gives rise to a Magnus force that opposes the interspecies vortex attraction.  
Beyond a critical
distance the Magnus force (Fig. \ref{Fig:cartoon}) becomes larger then the maximal pairing
force, and the vortices become unbound. In this region, the two
condensates and their vortices rotate together at the drive
frequency $\Omega_v$; the vortex densities in the two flavors are no longer
equal, but reflect themselves the mass ratio:
$\rho_v^1/m_1=\rho_v^2/m_2$.  Below we derive the characteristics and
conditions for the vortex locking state.

\emph{Energetics}.
The energy of weakly interacting Bose-Einstein condensates is 
well-described by
the Gross-Pitaevskii functional for the condensate wave function 
$\psi_{\alpha}$ \map{($\alpha = 1,2$)}
for the two atomic species \cite{pethick02}. We consider the situation
\map{in which} the two condensates are stirred at the same rate $\Omega_v$.
Transforming to the rotating frame, our problem becomes 
time independent.
The energy of the two-component system 
in the rotating frame is given by $E=E_1+E_2+E_{12}$, where
\begin{align} 
	E_{\alpha} &= \int d^2 r \left( \frac{\hbar^2}{2m_{\alpha}} |\nabla
\psi_{\alpha}|^2 + 
V_{\rm trap}({\bf r}) n_\alpha
\right. \notag \\
	&\qquad + \left. \frac{g_{\alpha}}{2} n_\alpha^2 - \hbar \Omega_v \;
\psi_{\alpha}^* 
\left( -i\frac{\partial}{\partial \varphi} \right)
	\psi_{\alpha} \right)
\end{align}
describes the energy for each atomic species \map{and} $g_{\alpha}$ is the intraspecies coupling \map{for bosons of flavor $\alpha$.}  The interspecies
interaction is given by
\begin{equation}
\label{Eq:interspecies}
	E_{12}=g_{12} \int d^2 r \map{\left[\right.}n_{1}({\bf r}) n_{2}({\bf r})\map{\left.\right]}\,\map{,}
\end{equation}
%
\map{where} $n_{\alpha}= |\psi_{\alpha}|^2$ is the density of flavor $\alpha$,
$V_{\rm trap}$ is the external trapping potential,
\map{and} the $z$-component of the angular momentum operator is
$L_z = -i \frac{\partial}{\partial \varphi}$
\map{(}where $\varphi$ is the azimuthal
coordinate\map{)}.  

The energy $E_{\alpha}$ of \map{one BEC component}
is
minimized via the nucleation of a vortex lattice rotating with the external
drive $\Omega_v$. In the following, we assume that the coherence length
$\xi_\alpha=\sqrt{\frac{\hbar^2}{2 m_\alpha g_{\alpha} n_0^{\alpha}}}$
is much shorter than the characteristic 
distance between
vortices. 
\map{E}ach vortex is then well\map{-}described by a small core region of size
$\xi_{\alpha}$, 
\map{at which} the superfluid density drops to zero
and its phase field
account\map{s} for the flow around the vortex. 
\map{T}he vortex core region gives
rise to a small constant energy, \map{and} we can account for the phase field by \map{writing}
$\psi_{\alpha}=\sqrt{n_{\alpha}} e^{i \theta_{\alpha}}$\map{,} 
\map{where} $\theta_{\alpha}$  
determine\map{s} a 
lattice of vortices 
\map{with} unit
winding
number at the positions $\{ {\bf r}_i^{\alpha}\}$.  We assume that we can write
$\theta_{\alpha}$ as a sum of the different vortex contributions
$\theta_{\alpha}=\theta_1^{\alpha} + \theta_2^{\alpha} + \dots$.  

With these assumptions, $E_\alpha$ can be written 
in terms of the positions of the vortices as 
\begin{align}
	E_{\alpha}= \frac{\hbar^2\pi}{m_{\alpha}}n_0^{\alpha} \sum_{i \ne j} 
\log \left(\frac{\xi_{\alpha}}{|{\bf r}_i^{\alpha}-{\bf r}_j^{\alpha}|}\right) 
+\hbar \Omega_v n_0^{\alpha} \pi \sum_i (r_i^{\alpha})^2\,, 
\label{Eq:Ealpha}
\end{align}
where we have dropped terms that do not depend on the positions of 
the vortices and have also neglected effects due to the 
nonuniform superfluid density \cite{sheehy04}.  The
first term in Eq.~(\ref{Eq:Ealpha}) 
is the usual logarithmic interaction between
vortices, and the second is the centripetal energy, reflecting the fact that
vortices toward the edge of the cloud carry less angular momentum
relative to the center of the cloud. In a single-component rotating
BEC\map{,} the balance of the two terms gives the Feynman condition 
(\ref{Eq:feynman}).   The
equations describe charged particles interacting in two dimensions with a
uniform background charge of opposite sign.

The energy $E_{12}$ arising from the interspecies interaction energy is
less straightforward to evaluate.  Unlike the intraspecies logarithmic
interaction, this nonuniversal interaction depends on the details of the
short-distance density variations around the vortex cores.  For instance,
in \cite{reijnders04} to study the interaction of a vortex with an
optical lattice, a step function having the width of the BEC coherence
length was taken. In this work we take a Gaussian 
depletion around the vortex core:
\begin{equation}
n_{\alpha}({\bf r})=n_0^{\alpha}(1-e^{-|{\bf r}-{\bf r}_0|^2/\xi_\alpha^2})
\label{Eq:gaussden}
\end{equation}
so that the system will be amenable to analytic treatment.
For a single vortex this depletion gives the correct behavior at short
distances, but not the long distance behavior, in which the density
due to a single isolated vortex heals as $\xi^2/r^2$. As we show in
the Appendix, the combined density variations due to the vortex lattice on
scales larger than the inter-vortex separation combines only to change the chemical potential (which is proportional to the density
correction), by 
$\frac{\hbar^2}{2m} \pi^2 \rho_v^2 r^2$, which just reflects the
kinetic energy associated with uniform rotation of the condensate. This can be shown to have a
negligible effect on the vortex pairs, which are the focus of this
work. Neglecting this piece of the density fluctuation is also consistent with neglecting 
the intraspecies core-core 
interactions which is a standard approximation \cite{fetter01}. 
Indeed, the short distance region is the relevant one for vortex locking;
two-species vortex pairs become unbound once their
separation is comparable to the coherence lengths, where the
approximate form we take for the density profiles is still valid.
Evaluating the interaction integral in Eq.~(\ref{Eq:interspecies}), 
and keeping only the
contributions due to the interactions between pairs of vortices between
different species, gives
\begin{equation}
	E_{12}=g_{12} n_0^1 n_0^2 \pi
\frac{\xi_1^2 \xi_2^2}
{\xi_1^2+\xi_2^2} 
\sum_{i  j} e^{-\frac{|{\bf r}_i^1-{\bf r}_j^2|^2}{\xi_1^2+ \xi_2^2}}\,.
\label{Eq:E12}
\end{equation}
We note that to prevent phase separation\map{,} the criterion
$|g_1 g_2| > |g_{12}|^2$ must be satisfied.
\map{Equations} (\ref{Eq:Ealpha}) and (\ref{Eq:E12}) now give the 
energetics of
the system purely in terms of the positions of the vortices.

\emph{Forces.}
An understanding of the locked phases can be
obtained by considering the forces acting on the vortices.
\map{Equation} \map{(}\ref{Eq:Ealpha}\map{)} leads to the 
well-known Magnus force acting on a
vortex of species $\alpha$ \map{\cite{donnelly91}:} 
\begin{equation}
	{\bf F}_{\rm mag}^{\alpha}= 2\pi \hbar  n_{0}^{\alpha} ({\bf
v}_{SF}^{\alpha} - {\bf v}_{v})
\times {\bf \map{\hat{\kappa}}}\map{\,,} 
	\label{Eq:Magnus}
\end{equation}
where ${\bf \map{\hat{\kappa}}}$ is a unit vector pointing out of the plane, 
$n_0^{\alpha}$ is the equilibrium superfluid density for 
species $\alpha$ (evaluated
away from the vortex core), 
${\bf v}_{SF}^{\alpha}=\frac{\hbar}{m_{\alpha}} \nabla \theta_{\alpha}$ 
is the superfluid velocity for species $\alpha$ (with the vortex
on which the force operates excluded from $\theta_{\alpha}$), and ${\bf v}_v$ 
is the velocity of the vortex.
On the other hand, the force arising from the energy 
in Eq.~(\ref{Eq:E12})  provides 
an attractive force between two vortices of different
species\map{.} 
\map{It} has the form
\begin{equation} 
	{\bf F}_{\rm{rstr}}^{\alpha}=-2\pi |g_{12}| n_0^{1}
n_{0}^2\frac{\xi_1^2\xi_2^2}{(\xi_1^2+\xi_2^2)^2}
e^{-\frac{{\bf d}^2}{\xi_1^2+\xi_2^2}} ~~{\bf d} \map{\,,}
\end{equation}
where ${\bf d}$ is the displacement vector between vortices.

Let us first briefly consider the unlocked case 
where the vortex interspecies interaction force is small. 
Since the force counteracting the Magnus force is too small, to bind
vortex pairs, we must have ${\bf F}_{\rm mag}=0$ for any isolated vortex\map{,} which implies \map{that} the
superfluid velocity must be the same as the vortex velocity\map{.} 
\map{That is,} the vortex lattice 
rotate\map{s} with the superfluid.  Thus\map{,} 
\map{because} the two
vortex lattices rotate at the same frequency, the two superfluids 
rotate
together at that frequency.  Accordingly, for this case, 
\map{the} 
vortex densities \map{are not equal}:
$\rho_v^1 = \frac{m_1}{m_2} \rho_v^2$.

We next consider the other extreme, in which the two vortex lattices are
locked. Our approach 
\map{is} to consider the forces acting on a bound
pair of vortices at distance $r$ from the center of the trap 
(see Fig.~\ref{Fig:cartoon}).  As
stated before, 
the Magnus force \map{for the locked state} 
\map{is} nonzero
\map{because} the superfluids 
\map{are} rotating at different rates.
Balancing the forces 
\map{gives} $F^{\alpha}_{\rm
mag}=F^{\alpha}_{\rm rstr}$.  
\map{Because} the restoring force acting on either
species 
\map{has} the same 
magnitude, we \map{obtain}
$F^{1}_{\rm mag}=F^{2}_{\rm mag}$. Noting that 
$v_\alpha= \Omega_{\alpha} r$  and $v_v= \Omega_{v} r$ (we are assuming
that $r$ is much larger than the distance between the two vortices) 
gives $n_0^1 (\Omega_v
- \Omega_1)= n_0^2 (\Omega_2 - \Omega_v)$.  This, along with the 
condition $m_1 \Omega_1= m_2 \Omega_2$ (from $\rho_v^1=\rho_v^2$) gives 
the \map{following} relation between
the angular velocities\map{:}
\begin{equation}
	\Omega_1= \frac{(n_0^1 + n_0^2)m_2}{m_1 n_0^2+m_2 n_0^1} \Omega_v < \Omega_v 
 < \Omega_2 = \frac{(n_0^1 + n_0^2)m_1}{m_1 n_0^2+m_2 n_0^1}  \Omega_v \,.
\end{equation}

Note that unlike the restoring force, the Magnus force grows linearly
with distance from the
center of the condensate.  Thus, at some critical distance
$r_c$, the pairs of vortices invariably become
unbound from each other. for radii $r>r_c$ the unlocked phase applies,
and after a short healing region, the two condensates rotate at the
same frequency. An expression for $r_c$ can be obtained by
equating the Magnus force
with the maximum possible value for the restoring force\map{:} 
\begin{equation}
r_c=\sqrt{\frac{1}{2e}} \frac{|g_{12}|}{\hbar \Omega_v} 
\frac{m_1 n_0^{2}+m_2 n_0^1}{m_1-m_2} \; \frac{\xi_1^2\xi_2^2}{(\xi_1^2+\xi_2^2)^{3/2}}\,.
\label{Eq:rc}
\end{equation}
Note that \map{(\ref{Eq:rc})}
diverge\map{s} 
\map{when the} masses \map{are equal}.
In addition, the bound pairs of vortices 
\map{are} pulled further
apart at increasing distances
from the center of the condensate.  
\map{The} interspecies vortex
separation $x_v$ 
\map{satisfies}
\begin{equation}
x_v e^{-\frac{x_v^2}{\xi_1^2+\xi_2^2}} = \frac{r}{r_c \sqrt{2e}} \; 
\sqrt{\xi_1^2+\xi_2^2} \map{\,,}
\end{equation}
which is valid for $r<r_c$. This 
introduce\map{s} a small correction to
the vortex density
and creat\map{es} a small shear in the motion of the two
condensates. 

For the vortex-locked state to be stable up to the critical radius
$r_c$, the superfluid velocities in the
rotating frame of the vortex lattice\map{,} \map{$|{\bf v}_{SF}^{\alpha}-{\bf v}_v|$}\map{,}
must not exceed the critical velocity of the superfluid.  Otherwise,
it would be possible to create elementary excitations from the flow
of the superfluid around the vortices.
For the two coupled superfluids, the Bogoliubov elementary excitations
are given by
\begin{equation}
	(\Omega_k)^2= \frac{1}{2} (E_{k}^1 + E_{k}^2)\pm
\frac{1}{2}\sqrt{(E_{k}^1 - E_{k}^2)^2+ 16 g_{12}^2 n_0^1 n_0^2 
\varepsilon_k^1 \varepsilon_k^2}\map{\,,}
\end{equation}
where 
\map{$E_{k}^{\alpha}=\sqrt{(\varepsilon_k^{\alpha})^2 + 
2 g_{\alpha} n_0^{\alpha}
\varepsilon_k^{\alpha}}$} and 
$\varepsilon_{k}^{\alpha}=\frac{\hbar^2 k^2}{2m_{\alpha}}$.  It is
then straightforward to compute the critical  velocity
$v_c = {\rm min}_k (\frac{\Omega_k}{\hbar k})$\map{; one obtains}
\begin{equation}
v_c = {\rm min}_{\alpha} \left\{  \sqrt{\frac{g_{\alpha}
    n_0^{\alpha}}{m_{\alpha}}}\right\}={\rm min}_{\alpha}
    \left\{\frac{\hbar}{\sqrt{2} m_{\alpha} \xi_{\alpha}}\right\}\map{\,.}
\end{equation}
\map{T}he superfluid velocity of 
species 1 or 2 in the vortex lattice frame 
evaluated at $r_c$ (where it is maxim\map{al}) is \map{given by} 
\begin{equation}
|{\bf v}_{SF}^{\{1,2\}}-{\bf v}_v| =
\frac{1}{\sqrt{2e}}\frac{|g_{12}|}{g_{\{2,1\}}}\frac{\hbar}{2m_{\{2,1\}}}
\frac{\xi_{\{1,2\}}^2}{(\xi_1^2+\xi_2^2)^{3/2}}\map{\,.}
\end{equation}
The condition \map{$|{\bf v}_{SF}^{\alpha}-{\bf v}_v|<v_c$} must be checked
so that the vortex-locked state is stable against 
\map{the creation of} elementary
excitations.  For instance,  it can be shown
that the system is stable against creating such elementary excitations
if the conditions $\frac{1}{10}\leq \frac{n_1^0}{n_2^0}\leq10$ 
and  $\frac{1}{10}\leq \frac{m_1}{m_2}\leq10$ are satisfied.

\begin{figure*}
\includegraphics[width=7in]{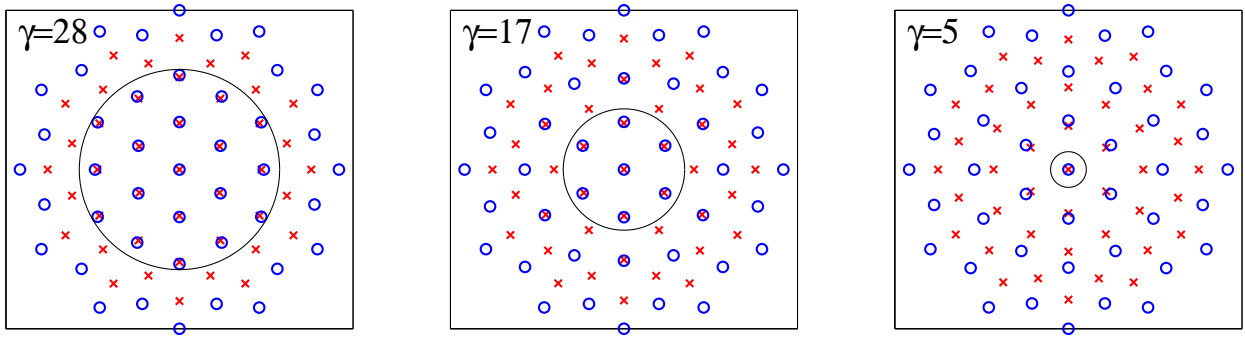}
\caption{Vortex lattice for two-component condensate
\map{(}with 43 vortices of each 
species\map{)}
for different values of
$\gamma=|g_{ab}| n_0/(\hbar \Omega_v)$.  Circles are shown 
for the theoretical
prediction for the critical radius Eq.~(\ref{Eq:rc2})
at which the vortices become unbound\map{.} 
}
\label{Fig:vlat}
\end{figure*}

\emph{Numerical Simulations.}  
Now that the expected types of phases have been discussed, we 
minimize the total energy $E=E_1+E_2+E_{12}$
as a function of the vortex positions given by Eq\map{s}.~(\ref{Eq:Ealpha}) and
(\ref{Eq:E12}).  
The ability to compute analytical expressions for the gradients of the
energy as a function
of the vortex positions 
$\nabla_{\lbrace {\bf r}_i \rbrace}E$
allow us to apply 
\map{the} steepest descent
method
for the minimization.  Specifically, 
starting with an initial configuration for the vortex
positions $\lbrace {\bf r}_i^{(0)} \rbrace$\map{,} we perform the
one-dimensional minimization of 
\begin{equation}
	E\left(\lbrace {\bf r}_i^{(n)} \rbrace - \lambda 
	\nabla_{\lbrace {\bf r}_i \rbrace} E \right)
\end{equation}
over $\lambda$\map{,} where $\nabla_{\lbrace {\bf r}_i \rbrace} E$ is 
evaluated at  $\lbrace {\bf r}_i^{n} \rbrace$.  
The new vortex positions are given by 
\begin{equation}
\lbrace {\bf r}_i^{(n+1)} \rbrace = 
\lbrace {\bf r}_i^{(n)} \rbrace - \lambda_{\rm min} 
\nabla_{\lbrace {\bf r}_i \rbrace} E\map{\,,}
\end{equation}
and the above procedure is repeated until \map{it} converge\map{s}.

To simplify the analysis, we restrict our attention to the
case in which the equilibrium densities
\map{and} healing lengths
of the two condensate components 
are equal: $n_0 \equiv n_0^1 = n_0^2,\,\xi\equiv \xi_1=\xi_2$.  
\map{Motivated by the example of a $^{133}$Cs-$^{87}$Rb condensate 
\cite{collision}, 
w}e fix the mass ratio
\map{to be} $m_1/m_2=1.5$. 
The vortex 
interaction strength is
parametrized by $\gamma=\frac{|g_{12}| n_0}{\hbar \Omega_v}$\map{,} which 
we vary while keeping  the quantities
$\frac{\hbar \pi}{\Omega_v m_{\alpha}} \frac{1}{\pi \xi^2}$ for
$\alpha=1,2$ fixed.  
\map{(T}he
total energy \map{has been scaled} by $\hbar \Omega_v \pi n_0 \xi^2$.\map{)}
We set the ratio of the ``average'' vortex lattice constant 
to the coherence length 
\map{${a_{\rm lat}}/{\xi}$}=10 (which is consistent with typical 
experiments).  
\map{W}e define $a_{\rm lat}$ by
$\frac{2}{\sqrt{3} a_{\rm lat}^2} = \frac{\map{\tilde{m}} 
\Omega_v}{\pi \hbar}$\map{,}  
where $\map{\tilde{m}}=\frac{2m_1m_2}{m_1+m_2}$\map{, and}
consider a system \map{with}
43 vortices of each species.
The results for such a calculation are shown in Fig.~\ref{Fig:vlat}.
\map{We also} plot
\map{our} prediction for the critical radius at which the
vortex pairs become unbound \map{[}see Eq.~(\ref{Eq:rc})\map{],} which 
for equal 
densities is 
\begin{equation}
r_c = \frac{\gamma}{4 \sqrt{e}} \frac{m_1+m_2}{m_1-m_2} \; \xi.
\label{Eq:rc2}
\end{equation}
\map{T}his prediction agrees quite well with 
\map{our} numerical results.

\emph{Experimental realization}.
A very promising candidate for the realization of these locked states 
\map{is a} 
$^{133}$Cs-$^{87}$Rb condensate \map{mixture} \cite{collision}\map{,}
\map{which} has a mass ratio of 
\map{about} $1.5$\map{.} 
\map{O}ne has exquisite control over the self-scattering length of 
Cesium  
\cite{cesium03}\map{, and a Cs-Rb} mixture is also expected to exhibit interspecies Feshbach resonances.  
This allows one to control the interaction $g_{12}$ over a wide range; such interspecies resonances
\map{have recently been identified for Li-Na \cite{stan04} and Rb-K \cite{inou04} mixtures.}



{\it Conclusions.}  In this paper\map{,} we described a
novel state of rotating interacting condensates with unequal masses in the
Thomas-Fermi regime. The possibility of locking the two individual 
vortex lattices \map{yields} a remarkable demonstration of the 
\map{nonintuitive} behavior
of superfluids: the two gasses, rather than equilibrat\map{ing} to the same
speed, prefer to move at different angular velocities \map{that are} inversely
proportional to their masses. The vortex-locking of the different-mass
condensates is also an example of synchronization: a phenomenon that is ubiquitous in physics,
biology, and other fields \cite{sync}. 
Already in single-mass mixtures, a rich variety
of vortex dynamics arises from the extra degrees-of-freedom, resulting in such
effects as the formation of square vortex lattice, as well as topologically 
nontrivial defects such as skyrmions or 
hedgehogs \cite{kasamatsu05a,kasamatsu05b}.  
To investigate these effects in the different-mass mixtures, 
as well as to better establish the
locked state we proposed in this manuscript, this system must be
numerically investigated by solving the appropriate dynamical Gross-Pitaevskii
equations.  Such numerical investigation will also allow one to find
the preferred lattice geometry of the locked vortex-lattice. 
Other directions for future study involve dynamical aspects
such as Tkachenko modes \cite{keceli06}) of the locked state, as well
as whether a similar state survives in the Landau regime of large
vortex density.


{\it Acknowledgements.}    
We thank the hospitality of the Kavli Institute for Theoretical Physics
where part of this work was completed.
We acknowledge support from the Sherman Fairchild Foundation
(RB), the Gordon and Betty Moore Foundation 
through Caltech's Center for the Physics of Information (MAP), and 
the National Science Foundation 
under Grant No. PHY05-51164 (RB, GR, HPB).  We \map{also} acknowledge useful
discussions with \map{Simon Cornish}, Michael Cross, Peter Engels, 
and Erich Mueller.


\section{Appendix}

In this appendix we discuss the change in the
superfluid density as a result of a vortex lattice, and its effect on
the validity of approximation (\ref{Eq:gaussden}) and the resulting
expression for the interspecies vortex attraction, Eq. (\ref{Eq:E12})
and (\ref{Eq:interspecies}).

First consider a single species which has the energy functional
\begin{equation}
E=\int d^2 r \left( \frac{\hbar^2}{2m} |\nabla \psi|^2  
+ V_{\rm trap} (r) |\psi|^2
+\frac{1}{2}g |\psi|^4 \right).
\end{equation}
We write $\psi = f e^{i\theta}$ where $f$ is real and $\theta$
contains information about the positions of the vortices as
$\theta = \sum_i \theta_i$.  By varying $f$ we obtain the equation
determining the superfluid density which minimizes $E$
\begin{equation}
-\frac{\hbar^2}{2m}\nabla^2 f + \frac{\hbar^2}{2m} |\nabla \theta|^2 f
+ V_{\rm trap} f + g f^3 = \mu f
\label{Eq:apen}
\end{equation}
for the particular vortex configuration.  First let us consider a single
vortex taken to be at the origin 
$\nabla \theta=\hat{z}\times\frac{\hat{r}}{r}$.
The long-distance
behavior $r \gg \xi$ is obtained from the 
Thomas-Fermi approximation (neglecting
the $\nabla^2 f$ in Eq.~\ref{Eq:apen}) and 
we obtain
\begin{equation}
f\approx \frac{\mu}{g}\left(1-\frac{\hbar^2}{2m\mu r^2}\right)=n_0\left(1-\frac{\xi^2}{r^2}\right)
\label{PW}
\end{equation}
where we have neglected the contribution from the trapping potential.
This implies that the suppression of the density is due to the
kinetic energy in the supercurrent, which counters the condensation
energy of the BEC. 

Eq. (\ref{PW}) seems to imply that at large distances from a vortex
core, the interspecies vortex-vortex interaction will include a
persistent power-law component, and die off only as $1/r^2$, rather
than as an exponential. But the observation that the power law decay
reflects the current-induced superfluid suppression, allows as to
ignore the power-law decay in a many-vortex situation, with the argument as follows.
Let us consider a vortex lattice, and evaluate $f$ at a position
which is several coherence lengths away from any vortex.  This allows
us to invoke the continuum approximation and write 
\begin{equation}
\nabla \theta (\bf r) = \hat{z} \times \sum_{i} 
\frac{{\bf r} - {\bf r}_i}{|{\bf r} - {\bf r}_i|^2} = \pi \rho_v
\hat{z} \times {\bf r}
\end{equation}
where $\rho_v$ is the density of the vortices.  
Each individual
contribution alone in this sum would yield a $\propto
\frac{1}{r^2}$ dependence in the density corrections, but the vector
sum of  the velocities of all vortices squared is not a simple sum of the
$1/r^2$ corrections. 
Inserting this into
the equation for the density profile one finds:
\begin{equation}
-\frac{\hbar^2}{2m}\nabla^2 f + \frac{\hbar^2}{2m} \pi^2 \rho_v^2 r^2 f
+ V_{\rm trap} f + g f^3 = \mu f.
\label{rs}
\end{equation}
Thus the combined vortex effect re-normalizes the
trapping potential and does not need to be explicitly taken into account.
To get an estimate for the magnitude of such a renormalization, one
can compare this term with the chemical potential
\begin{equation}
\frac{\frac{\hbar^2}{2m} \pi^2 \rho_v^2 r^2}{\mu} \sim 
\left(\frac{\xi}{a_{lat}}\frac{r}{a_{lat}}\right)^2
\end{equation}
where $a_{lat}$ is the vortex lattice
constant which is  small for typical experiments.

For a two-component BEC, the situation is similar for vortices which
are many coherence lengths away from each other.  On the other hand,
when the cores of the different types of overlap, their interaction
needs to be explicitly calculated, and the continuum approximation
cannot be used. This is the case for paired-vortex configurations.  Since the combined effect of
far-away vortices on a locked pair is small, the locking depends only on the
short distance density profile taken.
Had we used the 
step-function potential interaction between two vortices of 
Ref.~\onlinecite{reijnders04} the results would only differ
from the Gaussian depletion Eq.~\ref{Eq:gaussden} by small 
quantitative amounts. 


\end{document}